
\font\twelverm=cmr10 scaled 1200    \font\twelvei=cmmi10 scaled 1200
\font\twelvesy=cmsy10 scaled 1200   \font\twelveex=cmex10 scaled 1200
\font\twelvebf=cmbx10 scaled 1200   \font\twelvesl=cmsl10 scaled 1200
\font\twelvett=cmtt10 scaled 1200   \font\twelveit=cmti10 scaled 1200

\skewchar\twelvei='177   \skewchar\twelvesy='60


\def\twelvepoint{\normalbaselineskip=12.4pt
  \abovedisplayskip 12.4pt plus 3pt minus 9pt
  \belowdisplayskip 12.4pt plus 3pt minus 9pt
  \abovedisplayshortskip 0pt plus 3pt
  \belowdisplayshortskip 7.2pt plus 3pt minus 4pt
  \smallskipamount=3.6pt plus1.2pt minus1.2pt
  \medskipamount=7.2pt plus2.4pt minus2.4pt
  \bigskipamount=14.4pt plus4.8pt minus4.8pt
  \def\rm{\fam0\twelverm}          \def\it{\fam\itfam\twelveit}%
  \def\sl{\fam\slfam\twelvesl}     \def\bf{\fam\bffam\twelvebf}%
  \def\mit{\fam 1}                 \def\cal{\fam 2}%
  \def\tt{\twelvett}
  \textfont0=\twelverm   \scriptfont0=\tenrm   \scriptscriptfont0=\sevenrm
  \textfont1=\twelvei    \scriptfont1=\teni    \scriptscriptfont1=\seveni
  \textfont2=\twelvesy   \scriptfont2=\tensy   \scriptscriptfont2=\sevensy
  \textfont3=\twelveex   \scriptfont3=\twelveex  \scriptscriptfont3=\twelveex
  \textfont\itfam=\twelveit
  \textfont\slfam=\twelvesl
  \textfont\bffam=\twelvebf \scriptfont\bffam=\tenbf
  \scriptscriptfont\bffam=\sevenbf
  \normalbaselines\rm}

\def\beginlinemode{\endmode
  \begingroup\parskip=0pt \obeylines\def\endmode{\par\endgroup}}
\def\beginparmode{\endmode
  \begingroup \def\endmode{\par\endgroup}}
\let\endmode=\par
{\obeylines\gdef\
{}}
\def\singlespace{\baselineskip=\normalbaselineskip}
\def\oneandahalfspace{\baselineskip=\normalbaselineskip
  \multiply\baselineskip by 3 \divide\baselineskip by 2}
\def\doublespace{\baselineskip=\normalbaselineskip \multiply\baselineskip by 2}
\newcount\firstpageno
\firstpageno=2
\footline={\ifnum\pageno<\firstpageno{\hfil}\else{\hfil\twelverm\folio\hfil}\fi}
\let\rawfootnote=\footnote            
\def\footnote#1#2{{\rm\singlespace\parindent=0pt\rawfootnote{#1}{#2}}}
\def\raggedcenter{\leftskip=4em plus 12em \rightskip=\leftskip
  \parindent=0pt \parfillskip=0pt \spaceskip=.3333em \xspaceskip=.5em
  \pretolerance=9999 \tolerance=9999
  \hyphenpenalty=9999 \exhyphenpenalty=9999 }
\def\dateline{\centerline{\ifcase\month\or
  January\or February\or March\or April\or May\or June\or
  July\or August\or September\or October\or November\or December\fi
  \space\number\year}}
\hsize=6truein
\vsize=8.5truein

\parskip=\medskipamount
\twelvepoint         
\doublespace         
\overfullrule=0pt    
\def\title                      
  {\null\vskip 3pt plus 0.2fill
   \beginlinemode \doublespace \raggedcenter \bf}

\def\author                     
  {\vskip 3pt plus 0.2fill \beginlinemode
   \singlespace \raggedcenter}

\def\affil                      
  {\vskip 3pt plus 0.1fill \beginlinemode
   \oneandahalfspace \raggedcenter \sl}

\def\abstract                   
  {\vskip 3pt plus 0.3fill \beginparmode
   \doublespace \narrower ABSTRACT: }

\def\endtitlepage               
  {\endpage                     
   \body}

\def\body                       
  {\beginparmode}               

\def\refto#1{$^{#1}$}           

\def\references                 
   {\noindent{\bf REFERENCES}\medskip
   \beginparmode
   \frenchspacing \parindent=0pt \leftskip=0.5truecm
   \parskip=5pt plus 3pt \everypar{\hangindent=\parindent}}

\gdef\refis#1{\indent\hbox to 0pt{\hss#1.~}}    

\def\endreferences{\body}
\def\endpage                    
  {\vfill\eject}

\def\endpaper                   
  {\endmode\vfill\supereject}

\def\ref#1{Ref. #1}                     
\def\Ref#1{Ref. [#1]}                     

\def\cite#1{{#1}}
\def\[#1]{[\cite{#1}]}
\def\(#1){(\call{#1})}
\def\call#1{{#1}}\def\taghead#1{{#1}}
\def\frac#1#2{{\textstyle{#1 \over #2}}}

\def\sla{\raise.15ex\hbox{$/$}\kern-.57em}
\def\leaderfill{\leaders\hbox to 1em{\hss.\hss}\hfill}
\def\twiddle{\lower.9ex\rlap{$\kern-.1em\scriptstyle\sim$}}
\def\bigtwiddle{\lower1.ex\rlap{$\sim$}}
\def\gtwid{\mathrel{\raise.3ex\hbox{$>$\kern-.75em\lower1ex\hbox{$\sim$}}}}
\def\ltwid{\mathrel{\raise.3ex\hbox{$<$\kern-.75em\lower1ex\hbox{$\sim$}}}}
\def\square{\kern1pt\vbox{\hrule height 1.2pt\hbox{\vrule width 1.2pt\hskip 3pt
   \vbox{\vskip 6pt}\hskip 3pt\vrule width 0.6pt}\hrule height 0.6pt}\kern1pt}

\catcode`@=11
\newcount\r@fcount \r@fcount=0
\newcount\r@fcurr
\immediate\newwrite\reffile
\newif\ifr@ffile\r@ffilefalse
\def\w@rnwrite#1{\ifr@ffile\immediate\write\reffile{#1}\fi\message{#1}}

\def\writer@f#1>>{}
\def\referencefile{
\r@ffiletrue\immediate\openout\reffile=\jobname.ref%
  \def\writer@f##1>>{\ifr@ffile\immediate\write\reffile%
    {\noexpand\refis{##1} = \csname r@fnum##1\endcsname = %
     \expandafter\expandafter\expandafter\strip@t\expandafter%
     \meaning\csname r@ftext\csname r@fnum##1\endcsname\endcsname}\fi}%
  \def\strip@t##1>>{}}

\def\citeall#1{\xdef#1##1{#1{\noexpand\cite{##1}}}}
\def\cite#1{\each@rg\citer@nge{#1}}

\def\each@rg#1#2{{\let\thecsname=#1\expandafter\first@rg#2,\end,}}
\def\first@rg#1,{\thecsname{#1}\apply@rg}
\def\apply@rg#1,{\ifx\end#1\let\next=\relax%
\else,\thecsname{#1}\let\next=\apply@rg\fi\next}%

\def\citer@nge#1{\citedor@nge#1-\end-}
\def\citer@ngeat#1\end-{#1}
\def\citedor@nge#1-#2-{\ifx\end#2\r@featspace#1
  \else\citel@@p{#1}{#2}\citer@ngeat\fi}
\def\citel@@p#1#2{\ifnum#1>#2{\errmessage{Reference range #1-#2\space is bad.}
    \errhelp{If you cite a series of references by the notation M-N, then M and
    N must be integers, and N must be greater than or equal to M.}}\else%
{\count0=#1\count1=#2\advance\count1
by1\relax\expandafter\r@fcite\the\count0,%
  \loop\advance\count0 by1\relax
    \ifnum\count0<\count1,\expandafter\r@fcite\the\count0,%
  \repeat}\fi}

\def\r@featspace#1#2 {\r@fcite#1#2,}
\def\r@fcite#1,{\ifuncit@d{#1}
    \newr@f{#1}
    \expandafter\gdef\csname r@ftext\number\r@fcount\endcsname%
                      {\message{Reference #1 to be supplied.}
                       \writer@f#1>>#1 to be supplied.\par}
\fi
\csname r@fnum#1\endcsname}
\def\ifuncit@d#1{\expandafter\ifx\csname r@fnum#1\endcsname\relax}%
\def\newr@f#1{\global\advance\r@fcount by1%
      \expandafter\xdef\csname r@fnum#1\endcsname{\number\r@fcount}}

\let\r@fis=\refis
\def\refis#1#2#3\par{\ifuncit@d{#1}%

    \w@rnwrite{Reference #1=\number\r@fcount\space is not cited up to now.}\fi%
  \expandafter\gdef\csname r@ftext\csname r@fnum#1\endcsname\endcsname%
  {\writer@f#1>>#2#3\par}}

\def\ignoreuncited{
   \def\refis##1##2##3\par{\ifuncit@d{##1}%
     \else\expandafter\gdef\csname r@ftext\csname
r@fnum##1\endcsname\endcsname%
     {\writer@f##1>>##2##3\par}\fi}}

\def\r@ferr{\endreferences\errmessage{I was expecting to see
\noexpand\endreferences before now;  I have inserted it here.}}
\let\r@ferences=\references
\def\references{\r@ferences\def\endmode{\r@ferr\par\endgroup}}

\let\endr@ferences=\endreferences
\def\endreferences{\r@fcurr=0
   {\loop\ifnum\r@fcurr<\r@fcount
    \advance\r@fcurr by 1\relax\expandafter\r@fis\expandafter{\number\r@fcurr}%
    \csname r@ftext\number\r@fcurr\endcsname%
  \repeat}\gdef\r@ferr{}\endr@ferences}
\let\r@fend=\endpaper\gdef\endpaper{\ifr@ffile
\immediate\write16{Cross References written on []\jobname.REF.}\fi\r@fend}

\catcode`@=12
\citeall\refto\citeall\ref\citeall\Ref
\catcode`@=11
\newcount\tagnumber\tagnumber=0
\immediate\newwrite\eqnfile\newif\if@qnfile\@qnfilefalse
\def\write@qn#1{}\def\writenew@qn#1{}
\def\w@rnwrite#1{\write@qn{#1}\message{#1}}
\def\@rrwrite#1{\write@qn{#1}\errmessage{#1}}
\def\taghead#1{\gdef\t@ghead{#1}\global\tagnumber=0}
\def\t@ghead{}\expandafter\def\csname @qnnum-3\endcsname
  {{\t@ghead\advance\tagnumber by -3\relax\number\tagnumber}}
\expandafter\def\csname @qnnum-2\endcsname
  {{\t@ghead\advance\tagnumber by -2\relax\number\tagnumber}}
\expandafter\def\csname @qnnum-1\endcsname
  {{\t@ghead\advance\tagnumber by -1\relax\number\tagnumber}}
\expandafter\def\csname @qnnum0\endcsname
  {\t@ghead\number\tagnumber}
\expandafter\def\csname @qnnum+1\endcsname
  {{\t@ghead\advance\tagnumber by 1\relax\number\tagnumber}}
\expandafter\def\csname @qnnum+2\endcsname
  {{\t@ghead\advance\tagnumber by 2\relax\number\tagnumber}}
\expandafter\def\csname @qnnum+3\endcsname
  {{\t@ghead\advance\tagnumber by 3\relax\number\tagnumber}}
\def\equationfile{\@qnfiletrue\immediate\openout\eqnfile=\jobname.eqn%
  \def\write@qn##1{\if@qnfile\immediate\write\eqnfile{##1}\fi}
  \def\writenew@qn##1{\if@qnfile\immediate\write\eqnfile
    {\noexpand\tag{##1} = (\t@ghead\number\tagnumber)}\fi}}
\def\callall#1{\xdef#1##1{#1{\noexpand\call{##1}}}}
\def\call#1{\each@rg\callr@nge{#1}}
\def\each@rg#1#2{{\let\thecsname=#1\expandafter\first@rg#2,\end,}}
\def\first@rg#1,{\thecsname{#1}\apply@rg}
\def\apply@rg#1,{\ifx\end#1\let\next=\relax%
\else,\thecsname{#1}\let\next=\apply@rg\fi\next}
\def\callr@nge#1{\calldor@nge#1-\end-}\def\callr@ngeat#1\end-{#1}
\def\calldor@nge#1-#2-{\ifx\end#2\@qneatspace#1 %
  \else\calll@@p{#1}{#2}\callr@ngeat\fi}
\def\calll@@p#1#2{\ifnum#1>#2{\@rrwrite{Equation range #1-#2\space is bad.}
\errhelp{If you call a series of equations by the notation M-N, then M and
N must be integers, and N must be greater than or equal to M.}}\else%
{\count0=#1\count1=#2\advance\count1 by1\relax\expandafter\@qncall\the\count0,%
  \loop\advance\count0 by1\relax%
    \ifnum\count0<\count1,\expandafter\@qncall\the\count0,  \repeat}\fi}
\def\@qneatspace#1#2 {\@qncall#1#2,}
\def\@qncall#1,{\ifunc@lled{#1}{\def\next{#1}\ifx\next\empty\else
  \w@rnwrite{Equation number \noexpand\(>>#1<<) has not been defined yet.}
  >>#1<<\fi}\else\csname @qnnum#1\endcsname\fi}
\let\eqnono=\eqno\def\eqno(#1){\tag#1}\def\tag#1$${\eqnono(\displayt@g#1 )$$}
\def\aligntag#1\endaligntag  $${\gdef\tag##1\\{&(##1 )\cr}\eqalignno{#1\\}$$
  \gdef\tag##1$${\eqnono(\displayt@g##1 )$$}}
\def\eqalignno#1{\displ@y \tabskip\centering
  \halign to\displaywidth{\hfil$\displaystyle{##}$\tabskip\z@skip
    &$\displaystyle{{}##}$\hfil\tabskip\centering
    &\llap{$\displayt@gpar##$}\tabskip\z@skip\crcr
    #1\crcr}}
\def\displayt@gpar(#1){(\displayt@g#1 )}
\def\displayt@g#1 {\rm\ifunc@lled{#1}\global\advance\tagnumber by1
        {\def\next{#1}\ifx\next\empty\else\expandafter
        \xdef\csname @qnnum#1\endcsname{\t@ghead\number\tagnumber}\fi}%
  \writenew@qn{#1}\t@ghead\number\tagnumber\else
        {\edef\next{\t@ghead\number\tagnumber}%
        \expandafter\ifx\csname @qnnum#1\endcsname\next\else
        \w@rnwrite{Equation \noexpand\tag{#1} is a duplicate number.}\fi}%
  \csname @qnnum#1\endcsname\fi}
\def\ifunc@lled#1{\expandafter\ifx\csname @qnnum#1\endcsname\relax}
\let\@qnend=\end\gdef\end{\if@qnfile
\immediate\write16{Equation numbers written on []\jobname.EQN.}\fi\@qnend}
\catcode`@=12

\def\section#1#2{{\bf #1}{\bf #2}}
\title{DECOHERENCE IN QUANTUM BROWNIAN MOTION}
\vskip 1cm
\author{Juan Pablo Paz}
\vskip 0.5cm
\affil{Theoretical Astrophysics, Los Alamos National Laboratory,
Los Alamos, NM 87545}

\abstract{We examine the
dependence of decoherence on the spectral density of the
environment as well as on
the initial state of the system. We use two
simple examples to illustrate some important effects.}

\section{}{Introduction}

Decoherence plays a major role in the transition from quantum to classical and
has attracted much attention in recent years (see Zurek (1991)).
The analysis of this process may allow us to understand in detail the
mechanism that prevents observations of
some quantum systems in superpositions of
macroscopically distinguishable states. In the light
of new technologies it can also help
us to devise experiments to probe the fuzzy boundary between the
quantum and the classical world. The interaction with an external
environment is the mechanism responsible for the supression of quantum
interference effects. Therefore, there are several questions that arise
naturally:
 How dependent on the environment decoherence is?  What are the time scales
involved in this process? How are some preferred states of the system
dynamically chosen? In this paper we will report on recent work where some of
these questions are addressed.

As a first point, let us clarify what we mean here by decoherence.
Within the Gell--Mann and Hartle version of the consistent histories
formulation of quantum mechanics, based on earlier work by Griffiths
and Omn\`es,  (see contributions in this proccedings) the term decoherence is
used in place of ``consistency'' which is the
condition that, if satisfied, allows us to assign probabilities to members
of sets of coarse grained histories of a closed system. The Decoherence
Functional is the basic diagnostic tool used in this framework. On the other
hand, in previous works originated in quantum measurement theory, a
different notion of decoherence was used. Measurement devices are always open
systems that interact with external environments. This interaction dynamically
selects a preferred set of states of the apparatus, the so--called pointer
basis.
This is, in some sense, the set of the most stable states: if the aparatus is
prepared in a pointer state, the interaction with the environment has a
minimal effect and almost no predictive power is lost. On the contrary, if the
initial state is a superposition of pointer states, the interaction with the
environment induces correlations and the state of the system tends to evolve
into a mixture of pointer states. This process was called decoherence and
this is the sense in which we will use this
word here. Within this context,
there are several important issues that require further attention.
The most important one seems to be the definition of an appropriate
measure of stability that may be used to
determine the pointer states (see Zurek's
contribution in this conference). We will study
a model describing a particle interacting with an environment formed by a
collection of harmonic
oscillators. In this case the pointer states of the particle seem to be
closely
related to coherent states and decoherence is the process
that supresses interference between coherent states (our diagnostic tool will
be described later). The action of the model is the following:
$$
S[x,q]=
 \int\limits_0^tds\Biggl[{1\over 2}M\Bigl(\dot x^2-\Omega_0^2 x^2\Bigr) +
     \sum_n
{1\over 2}m_n(\dot q_n^2- \omega^2_nq_n^2) -
     \sum_n C_nxq_n\Biggr]\eqno(1)
$$
We will assume that there are no initial correlations between the system and
the environment (i.e. the initial density matrix factorizes) and that the
initial state of the environment is in thermal equilibrium at temperature $T$.

In Section 2 we will describe generic features of the evolution of a
system
interacting with a general environment. We will study in detail the case in
which the initial state is a superposition of two coherent states. In Section 3
we will illustrate the fact that decoherence strongly depends on some
properties of the environment. We will also illustrate in what sense position
is a
preferred observable in the model (where position eigenstates are not
pointer states). In the Appendix we outline a simple derivation of the master
equation for a general environment.

\section{}{General Properties of the Reduced Dynamics}

Due to the interaction with the environment, the evolution of the system is
non--unitary since it is affected by a stochastic noise and a
``dissipative'' force (the word dissipation is used here in a rather vague
sense). Noise and dissipation, are entirely determined by two properties of the
environment: the spectral density $I(\omega)$ and the initial temperature. The
spectral density, defined as
$I(\omega)=\sum_n\delta(\omega-\omega_n)C_n^2/2m_n\omega_n$,
characterizes the number density of oscillators in the environment and the
strength of their coupling with the system. Therefore, in order to
analyze how decoherence depends on the environment we can study how this
process changes when varying $I(\omega)$ and the temperature since these are
the only two environmental properties ``seen'' by the system.

Luckily enough, the
reduced dynamics has some very general features that are entirely independent
of the spectral density and the temperature. One of the most striking and
important ones is the fact that the reduced density matrix {\it always}
satisfies a master equation that can be written as follows (we use $\hbar =1$):
$$
\eqalign{ i~\partial_t\rho_r(x,x',t)=&
<x|\bigl[H_{ren}(t),\rho_r\bigr]|x'>
 -i\gamma(t)(x-x')({\partial_x}-{\partial_{x'}})~ \rho_r(x,x',t)~ \cr
& -i~D(t)(x-x')^2~ \rho_r(x,x',t)~ + f(t)(x-x')({\partial_x}
  +{\partial_{x'}}) ~\rho_r(x,x',t) \cr}\eqno(2)
$$

\noindent This equation depends on the spectral density and the temperature
only through the coefficients appearing in the right hand side: the physical
frequency entering in $H_{ren}(t)$, the friction coefficient $\gamma (t)$ and
the diffusion coefficients $D(t)$ and $f(t)$ are time dependent functions that
vanish initially and depend on the environment in a fairly complicated way.

The validity of (2) for a general environment at arbitrary temperature has been
recently demonstrated by Hu et al. (1992) and is an interesting discovery that
generalizes previous results concerning the nature of the master equation (see
Unruh and Zurek (1989), Caldeira and Leggett (1985), Haake and Reiboldt
(1985)).
The result is also surprising since a general environment generates  a
non--Markovian evolution for which one expects highly nonlocal integral kernels
in the master equation. However, for this model the non--Markovian effects can
be fully encoded in the time dependence of the coefficients. The interested
reader can find a simple proof of equation (2) (the simplest I could think
off) in Appendix 1.

Equation (2) is a very useful tool to study generic properties of the evolution
and can be exactly solved for some simple initial conditions. To study
decoherence we will consider the following initial superposition of coherent
states:
$$
\Psi(x,t=0)= N \exp(-{{(x- L_0)^2}\over{2\delta^2}}+ iP_0 x)+ N \exp(-{{(x+
L_0)^2}\over{2\delta^2}}- iP_0 x)\eqno(3)
$$
where $N$ is a constant. In this case it is possible to solve the master
equation and show that the Wigner function constructed from the reduced density
matrix is:
$$
W(x,p,t) = W_1(x,p,t) + W_2(x,p,t) + W_{int}(x,p,t) \eqno(4)
$$
where
$${\eqalign{
W_{1\atop 2}(x,p,t) &= {{\bar N^2}}~{{\delta_2}\over{\delta_1}}
\exp\Bigl(-{{(x\mp x_c)^2}\over{\delta_1^2}}-\delta_2^2(p\mp p_c-\beta (x\mp
x_c))^2\Bigr)\cr
W_{int}(x,p,t) &= 2 {{\bar N^2}}~{{\delta_2}\over{\delta_1}}
\exp\bigl(-{{x^2}\over{\delta_1^2}}-\delta_2^2(p-\beta x)^2\bigr)
\cos\bigl(\phi_p p + (\phi_x-\beta \phi_p) x\bigr)~\exp(-A_{int})\cr}}\eqno(5)
$$
The functions $x_c(t),~P_c(t),~\delta_{1\atop 2}(t),~\beta(t)~,\phi_x(t),
\phi_p(t)$ and $A_{int}(t)$ depend on the environment (and on the constants
$L_0,~P_0$ and $\delta$ that appear in (3)) in a rather complicated way. For
the sake of brevity, we will not discuss  here the behavior of all these
functions
(see Paz et al (1992) for details) but concentrate on $A_{int}$ which is the
only one relevant for decoherence. Thus,
to quantify the importance of interference at a given time we will use the peak
to peak ratio between the interference and the direct terms in the Wigner
function, a quantity closely related to $A_{int}$:
$$
\exp(-A_{int})={1\over 2}
{{W_{int}(x,p)|_{peak}}\over
{\Bigl(W_1(x,p)|_{peak}W_2(x,p)|_{peak}\Bigr)^{1/2}}}\eqno(6)
$$
As the two initial wave packets
have a finite overlap, the above function satisfies
$A_{int}\leq\delta^2P_0^2+L_0^2/\delta^2$. The
system decoheres when $A_{int}$ irreversibly grows
to a value that is large with respect to unity (which can only occur if
the initial peaks are well separated, i.e. $delta^2P_0^2+L_0^2/\delta^2>>1$).

To analyze how the evolution of $A_{int}$ is affected by the environment it is
convenient to use the master equation (2) to show the following identity:
$$
\dot A_{int} = D(t) \phi_p^2 - 2 f(t) \phi_p (\phi_x -\beta \phi_p)\eqno(7)
$$
The first term carries the effect of normal diffusion and always produces
decoherence since increases the value of  $A_{int}$. On the contrary, the sign
of the second term in (7) may vary in time depending upon the relation between
$\phi_p$ and $\phi_x$.
Equation (7) can be approximately solved if one neglects the
anomalous diffusion and considers $D(t)$ as a constant, two
conditions met by an ohmic environment in the high tempearture
regime (see next section). In this case it can be shown that,
for an initial state with $P_0=0$,
$A_{int}(t)\simeq4L_0^2D{t}/({1+4D\delta^2t})$ and that
the ``decoherence rate''
is $\Gamma_{dec}=4L_0^2D\simeq8L_0^2m\gamma_0k_BT$
(see Paz et al. (1992)). However, this
solution is no longer valid when one moves away from the ohmic environment in
the high temperature regime or when considers more general initial
states. In fact, the behavior of
$A_{int}$ strongly depends on the initial conditions
(that enter into (7) through the functions $\phi_x$ and $\phi_p$ whose initial
data are $\phi_x=P_0,~\phi_p=L_0$) and
decoherence will be drastically different in the case $L_0=0,~P_0\neq 0$ where
the two initial gaussian are spatially separated than when  $P_0=0$ and
$L_0\neq0$ (where the coherent states are separated in momentum).

\section{} {Decoherence and the environment.}

A wide and interesting class of environments is defined by a spectral density
of the form
$I(\omega)={{2m\gamma_0}\over{\pi}}{{\omega^n}\over{\Lambda^{n-1}}}
\exp{(-{{\omega^2}\over{\Lambda^2}})}$
where $\Lambda$ is a high frequency cutoff and $n$ is an index that
characterizes different environments. We will consider two examples: $n=1$
which is the largely studied ohmic environment
(Caldeira and Leggett (1985)) and $n=3$ which is a supra--ohmic
environment used to model the interaction between defects and phonons in
metals, Grabert et al (1988), and also to mimic the interaction
between a charge and its own electromagnetic field, Barone and Caldeira (1991).

Using these two environments we want to illustrate how strongly decoherence
depends on the spectral density. It can be shown that the process is
much more inefficient in the supraohmic than in the ohmic case because the
final value of the diffusion coefficient $D(t)$
is much smaller in the former than in the latter environment (as
$n=3$ corresponds to a bath
of oscillators with an infrared sector substantially weaker than $n=1$,
the dissipative and diffusive effects are expected to be weaker). The time
dependence of the diffusion
coefficient for these two environments has been described
by Hu et al (1992) and has a rather generic feature: $D(t)$
vanishes initially and develops a very strong peak in a
time scale of the order of the collision time $\tau_\Lambda = 1/\Lambda$. Its
value after the initial peak is
$D(\tau_\Lambda)\simeq m\gamma_0\Lambda$, approximately the same for all
environments. After this initial cutoff dominated regime, $D(t)$
approaches (in a dynamical time scale) an asymptotic value
that depends on the environment (in the high temperature regime,
$D(t)\rightarrow 2m\gamma_0k_BT$ for $n=1$ while vanishes for $n=3$).
Thus, there is no generic long time behavior but a quite universal short time
regime.
One may thus wonder if this general initial behavior produces a rather
universal decoherence. We will argue here that this is not the case.
The impact of the initial peak has been analyzed in
detail (see Unruh and Zurek (1989)) and it was shown that in some cases may
completely wipe out the interference effects. However,
the physical significance of the decoherence produced by the initial peak is
rather questionable since this jolt is certainly related to the initial
conditions that do not contain correlations between the system and its
environment. In fact, such correlations are likely to wash out the initial
peak, Grabert et al (1988).

To discredit even more the role of the initial peak on decoherence we
would like to
point out that its effect can be made
completely inocuous by appropriately choosing the
initial conditions for the system. This is well illustrated by
the supra--ohmic environment where the asymptotic value of the
diffusion coefficient is too small to produce decoherence and all the
effect, if any, should come from the initial peak. In Figure 1 we plotted
$A_{int}$
for the ohmic and supraohmic environments.
We considered an harmonic oscillator with renormalized frequency $\Omega_r$ and
fixed $\gamma_0=0.3\Omega_r$, $\Lambda=500\Omega_r$ and $k_BT=25000\Omega_r$
(high temperature regime).

We can notice that in the ohmic environment decoherence is very
fast. For the initial condition I ($L_0=3\delta, P_0=0$)
it takes place in a time of the order of
$\tau_\Lambda$ while for condition II ($L_0=0, P_0=3/\delta$) it requires
a time that is also much smaller than $\Omega_r^{-1}$.
On the other hand, in the supraohmic environment of
Figure 1.b decoherence goes as in the ohmic case for condition I while no net
decoherence is achieved for condition II. In this case the initial
growth of $A_{int}$ is followed by a plateau and a decreasing regime during
which coherence is recovered! The reason
for the drastic difference between the fate of conditions I and II in the
supraohmic environment is clear:
decoherence can only be produced by
the initial peak but the interaction between the system and the environment is
initially effective only
if the two coherent states are spatially separated. The non monotonic
behavior of $A_{int}$ seen in curve (II) of Figure
1.b is  due to the anomalous diffusion
that cannot produce any net decoherence since the sign
of the second term in the r.h.s. of (7) changes with time.

The above example not only illustrates the strong dependence of decoherence
on the spectral density but also clarifies in what sense
position is an observable that is preferred by the interaction. In fact, in
general, coherent
states that are spatially separated decohere much faster than those
separated only in momentum (see Paz et al (1992) for more details).

\section{}{Acknowledgements:} I would like to thank the organizers for giving
me the opportunity to attend such an exciting meetting. I also want to thank
B.L.
Hu, W Zurek and S. Habib for many interesting conversations.

\section{}{Appendix: Derivation of the Master Equation}

I outline here a simple derivation of the master equation (2) based on the
properties of the evolution operator of the reduced density matrix.   This
propagator, which we denote as $J(t,t_0)$ and is defined so as to satisfy
$\rho_{red}(t)=J(t,t_0)\rho_{red}(t_0)$, has a path integral representation of
the following form:
$$
J(x,x',t~|~x_0,x'_0,t_0)
=\int\limits_{x_0}^{x}D\tilde x \int\limits_{x'_0}^{x'}D\tilde x'~
 \exp{i\over\hbar}\Bigl\{S[\tilde x]-S[\tilde x']\Bigr\}~F[\tilde x,\tilde
x']\eqno(A.1)
$$
where $F(x,x')$ is the Feynman--Vernon influence functional that arises due to
the integration of the environment variables. For the model we are considering,
this functional is well known and can be written as (see Grabert et al (1988)):
$$
i~\ln(F[x,y])=\int\limits_0^tds\int\limits_0^sds'
   (x-y)(s)\Bigl[\eta(s-s')(x+y)(s')-i\nu(s-s')(x-y)(s')\Bigr]
$$
where $\nu(s)$ and $\eta(s)$ are the noise and dissipation kernels defined in
terms of the spectral density:
$$
\eta(s)=-\int_0^\infty d\omega I(\omega) \sin(\omega
s),\qquad\nu(s)=\int_0^\infty d\omega I(\omega) \coth({\omega\over{2k_BT}})
\cos(\omega s)
$$
As the integrand of (A.1) is gaussian, the integral can be exactly computed and
the result is (written in terms of the variables $\xi=x-x'$, $X=x+x'$):
$$\eqalign{
J(X,\xi,t;X_0,\xi_0,t_0) = {b_3\over{2\pi}}~&\exp(i b_1 X\xi +i b_2X_0\xi -i
b_3 X\xi_0 -i b_4 X_0\xi_0)\times\cr
&\times\exp(-a_{11}\xi^2-a_{12}\xi\xi_0-a_{22}\xi_0^2)\cr}\eqno(A.2)
$$
where the functions $b_k(t)$ and $a_{ij}(t)$ depend on the environment and can
be constructed in terms of solutions to the equation:
$$
\ddot u(s) + \Omega_0^2 u(s) + 2\int_0^sds' \eta(s-s')~u(s') =0\eqno(A.3)
$$
Thus, if $u_1$ and $u_2$ are two solutions of (A.3) that satisfy the
boundary conditions $u_1(0)=u_2(t)=1$ and $u_1(t)=u_2(0)=0$ we can write:

$$
\eqalign{
2~b_1(t) = & \dot u_2(t), \quad 2~b_3(t) = \dot u_2(0), \quad
2~b_2(t) =  \dot u_1(t), \quad 2~b_4(t) = \dot u_1(0)\cr
a_{ij}(t)= & {1\over{1+\delta_{ij}}}\int_0^t\int_0^t ds~ds'
u_i(s)~u_j(s')~\nu(s-s')}\eqno(A.4)
$$

The derivation of the master equation can be done as follows
by simply using equations
(A.2) and (A.4): Let us take the time derivative of (A.2) and write
$$
\dot J(t,t_0)=\Bigl[{\dot b_3\over{b_3}}+ i\dot b_1 X\xi + i \dot b_2X_0\xi -i
\dot b_3 X\xi_0 -i \dot b_4 X_0\xi_0 -\dot a_{11}\xi^2-\dot a_{12}\xi\xi_0-\dot
a_{22}\xi_0^2\Bigr] J(t,t_0)\eqno(A.5)
$$
If we multiply (A.5) by an arbitrary initial density matrix and
integrate over the initial coordinates $\xi_0$ and $X_0$, we will obtain an
equation whose left hand side is $\dot\rho_r(x,x',t)$. In the right hand side
we will find
terms proportional to $\rho(x,x',t)$ that look like some of the
ones appearing in the right hand side of equation (2). The only potentially
problematic terms are the ones that in (A.5) are proportional to the initial
coordinates $\xi_0$ and $X_0$. However, their contribution can be easily shown
to be local by realizing that the propagator $J(t,t_0)$ satisfies:
$$
\eqalign{\xi_0 J(X,\xi,t;X_0,\xi_0,t_0) = & \bigl({{b_1}\over{b_3}}\xi +
{i\over {b_3}}\partial_X\bigr) J(X,\xi,t;X_0,\xi_0,t_0)\cr
X_0 J(X,\xi,t;X_0,\xi_0,t_0) = & \bigl(- X {{b_1}\over{b_2}} - {i\over{b_2}}
\partial_{\xi} - i({{2 a_{11}}\over{b_2}}+{{a_{12}b_1}\over{b_2b_3}})\xi
+{{a_{12}}\over{b_3b_2}}\partial_X\bigr) J(X,\xi,t;X_0,\xi_0,t_0)\cr}
$$
Using these equations, the right hand side of (A.5) can
be written in terms of the final coordinates $X$ and $\xi$ and the
derivatives with respect to them, which implies that the
master equation is local. To show that this equation is given by (2),
we just have to demonstrate that the coefficients associated to
terms like $\partial^2_X$ or $X\partial_X$ cancel and this can be done by
exploiting some general properties of the coefficients $b_k$ and $a_{ij}$ that
follow directly from their definition in (A.4). In fact, using relations such
as $\dot a_{22}=-\dot b_4 a_{12}/b_2$ (whose proof we omit), it is possible to
show that the coefficients of equation (2) are:
$$
\eqalign{
\Omega^2_{ren}(t)&=2({{\dot b_2b_1}\over {b_2}}-\dot b_1);\qquad
\gamma(t)=-(b_1+{{\dot b_2}\over{2b_2}})\cr
D(t)&=\dot a_{11}-4a_{11}b_1+\dot a_{12}{{b_1}\over{b_3}}-{{\dot
b_2}\over{b_2}}(2a_{11}+a_{12}{{b_1}\over{b_3}})\cr
2f(t)&={{\dot a_{12}}\over {b_3}}-{{\dot b_2 a_{12}}\over{b_2b_3}}-4a_{11}\cr
}
$$

\section {}{References}

\noindent Barone, P.M.V.B. and Caldeira, A.O. (1991) Phys. Rev. {\bf A 43}, 57

\noindent Caldeira, A.O. and  Leggett, A.J. (1985) Phys. Rev. {\bf A 31}, 1059

\noindent Grabert, H., Schramm, P. and Ingold, G. (1988) Phys. Rep. {\bf 168},
115

\noindent Haake, F and Reibold, R. (1985), Phys Rev {\bf A 32}, 2462

\noindent Hu B.L., Paz, J.P. and Zhang, Y. (1992) Phys Rev {\bf D45}, 2843

\noindent Paz, J.P., Habib, S.. and Zurek, W.H., (1992), Los Alamos Preprint

\noindent Unruh, W.G. and Zurek, W.H. (1989), Phys. Rev. {\bf D40}, 1071

\noindent Zurek, W.H. (1991), Physics Today {\bf 44}, 33

\vfill
\eject
\section{}{Figure Captions}

Figure 1: The degree of decoherence  $A_{int}$
is plotted as a function of time (which is measured
in units of $\Omega_r^{-1}$) for the
ohmic (a) and supra--ohmic (b) environments in the high temperature regime.
Curve (I) corresponds to an initial condition where the initial coherent states
are spatially separated ($L_0=3\delta,~P_0=0$) while for curve (II)
the initial separation is in momentum ($L_0=0,~P_0=3/\delta$).

\vfill
\eject
\section{}{Questions}

\noindent{\bf Unruh:} Your master equation is local in time. For an arbitrary
spectral density I would strongly expect that the equations are
strongly non local in time. Why aren't yours?

\noindent{\bf Paz:}  There are two observations one can intuitively make
for the model described by equation (1). On the
one hand we expect that a general
environment will produce non--Markovian effects
and that the master equation will be non local in
time. On the other
hand, the (reduced) evolution
operator must be gaussian since
the problem is linear. The crucial observation is that if one admits a
gaussian evolution operator, the master
equation is always local provided the matrix  mixing ``old'' and
``new'' coordinates in the propagator can be inverted. Taking this into
account, it is surprising to
me that the existence of a local master equation has not been noticed until
so recently.

\noindent{\bf Morikawa:} Why do you get a local coefficient $\gamma (t)$?

\noindent{\bf Paz:} I refer to the answer I gave to Prof. Unruh's question.

\end